\documentclass[twoside,leqno,twocolumn]{article}

\usepackage[letterpaper]{geometry}

\usepackage{ltexpprt}
\usepackage{amsmath}
\usepackage{hyperref}
\usepackage{color}
\usepackage{graphicx}

\begin{document}

\newcommand\relatedversion{}
\newcommand{\ve}{\varepsilon}

\title{\Large Optimizing Districting Plans to Maximize Majority-Minority Districts \\via IPs and Local Search\relatedversion}

\author{Daniel Brous\thanks{Cornell University.}
\and David Shmoys\thanks{Cornell University.}
}

\maketitle







\begin{abstract} \small\baselineskip=9pt
In redistricting litigation, effective enforcement of the Voting Rights Act has often involved providing the court with districting plans that display a larger number of majority-minority districts than the current proposal (as was true, for example, in what followed Allen v. Milligan concerning the congressional districting plan for Alabama in 2023). Recent work by Cannon et al. proposed a heuristic algorithm for generating plans to optimize majority-minority districts, which they called {\it short bursts}; that algorithm relies on a sophisticated random walk over the space of all plans, transitioning in bursts, where the initial plan for each burst is the most successful plan from the previous burst. We propose a method based on integer programming, where we build upon another previous work, the stochastic hierarchical partitioning algorithm, which heuristically generates a robust set of potential districts (viewed as columns in a standard set partitioning formulation); that approach was designed to optimize a different notion of fairness across a statewide plan. We design a new column generation algorithm to find plans via integer programming that outperforms short bursts on multiple data sets in generating statewide plans with significantly more majority-minority districts. These results also rely on a new local re-optimization algorithm to iteratively improve on any baseline solution, as well as an algorithm to increase the compactness of districts in plans generated (without impacting the number of majority-minority districts).
\end{abstract}

\section{Introduction.}

In the landmark case Thornburg v. Gingles (1986), the United States Supreme Court laid out a set of conditions on which to frame Voting Rights Act litigation concerning the racial gerrymandering of a particular minority group, with one of those being that ``...the minority group must be able to demonstrate that it is sufficiently large and geographically compact to constitute a majority in a single-member district'' \cite{thornburg}.  In bringing a lawsuit, condition is typically demonstrated by showing that it is possible to create a districting plan with more majority-minority districts than the current enacted plan \cite{negron, chestnut, johnson, dwight, milligan}. Due to the prevalence of this litigation tactic, it is important to provide algorithmic tools that construct (statewide) districting plans that maximize (or nearly maximize) the number of majority-minority districts that a single-member districting plan can support. Hence, we study the following computational problem: given a specified targeted subpopulation, find a plan that maximizes the number of districts in which that subpopulation constitutes a majority.

A significant challenge in understanding and exploring the space of possible districting plans is the sheer number of plans that are feasible. The exact requirements that districting plans must satisfy vary slightly state by state, but in each case a plan is created by taking a geographical partition of a region provided by the US Census Bureau and grouping the components into contiguous districts that have roughly equal populations. These census geographical partitions can break up states into hundreds of thousands of pieces, and so even for a small number of districts, the number of feasible plans makes complete enumeration of this space practically infeasible. Furthermore, even if the number of majority-minority districts is the sole objective, determining this optimal value is intractable, and is NP-hard with respect to any reasonable partisan fairness objective function \cite{puppe2008computational, kueng2018fair, chatterjee2019partisan}. Consequently, optimization approaches commonly involve integer programming techniques or some form of local search to find reasonably good solutions.

The literature on algorithmic approaches to redistricting is quite extensive, and an excellent place to start is the recent book edited by Duchin and Walch \cite{duchin2022political}, and in particular, its chapters \cite{becker2022redistricting} and \cite{deford2022random}. The two primary streams of work in this area are methods that are based on integer programming (and related optimization methods), and algorithms that are based on a Markov Chain Monte Carlo (MCMC) approach to exploring the space of statewide districting plans, originally as a means for gaining a probabilistic understanding of the distribution of all such legal plans. The former approach dates back over 50 years, starting with formative papers of Hess et al. \cite{hess1965nonpartisan} and Garfinkel \& Nemhauser \cite{garfinkel1970optimal}; the formative moment in the development of the MCMC approach was the Recombination Markov Chain (commonly referred to as ReCom) \cite{deford2019recombination}.

The state of the art to maximizing majority-minority districts based on the MCMC approach is the short bursts \cite{cannon2022votingrightsmarkovchains} algorithm, which builds on ReCom, by running in bursts of a few steps at a time. Each burst starts with the plan from the previous burst that had the most majority-minority districts. ReCom is agnostic to the number of majority-minority districts, and while this might lead to more compact districts, it might also limit the capability of short bursts to achieve near-optimal solutions.

Integer programming has also been used to explore various districting optimization scenarios, including maximizing majority-minority districts. It has recently been applied in the cases of preserving indigenous representation \cite{arredondo2021mathematical} and maximizing majority Black representation \cite{kroger2024bounding}, achieving good solutions on a small scale. A more scalable approach to optimizing districts using integer programming is the statistical hierarchical partitioning algorithm \cite{fairmandering}, which builds a large ensemble by recursively splitting a region into composable pieces using a partitioning integer program that can then be optimized over to find a plan that maximizes any of a large family of fairness objectives. Quite recently, a paper of Belotti, Buchanan, and Ezazipour \cite{belotti2024political} has also applied mixed integer second order cone programming (MISOCP) to attack this problem as well. Their approach finds districtings with higher average district compactness than our approach and more majority Black districts than short bursts. Our approach, however, finds more majority Black districts than theirs, as we did not directly ensure that our plans satisfy all Gingles conditions (the conditions that Thornburg v. Gingles outlined). Recent work of Kroger et al. \cite{kroger2024bounding} also provides an optimization framework for upper bounding the maximum number of majority-minority districts that are feasible.

In this paper, we adapt the methodology of the stochastic hierarchical partitioning algorithm \cite{fairmandering} to design a scalable integer programming approach which outperforms short bursts by optimizing for majority-minority districts both on a local and global scale, finding significantly more majority-minority districts in multiple states and for multiple minority groups. We also design a new local reoptimization algorithm which uses both integer programming and local search to improve on any baseline solution, as well as an algorithm to increase the compactness of districts in plans generated by either of these algorithms.

In broad outline, the stochastic hierarchical partitioning algorithm relies on randomization and an integer programming partitioning algorithm to recursively partition a region into subregions along with its allotment of representatives, producing multiple such partitions, until the base of the recursion where the resulting subregion has exactly one allocated representative. This provides a robust set of potential districts, and then a set partitioning integer program is solved that can optimize over any (piecewise) linear function of metrics captured at the potential district level. Although the previous application of this approach was geared towards finding statewide plans that optimize either the expected proportionality or efficiency gap (where the expectation is with respect to a stochastic model of recent historical voting patterns, precinct by precinct), one could clearly similarly optimize the number of majority-minority districts. This direct translation provided relatively weak results (barely matching those of the short bursts algorithm). However, by taking full advantage of the strength of the partitioning IP, we shall demonstrate that this framework of generating a rich portfolio potential districts that can be recombined together constitutes a powerful method for generating statewide districting plans.

In our experiments, we consider the data sets reported on by Cannon et al. \cite{cannon2022votingrightsmarkovchains}, and in each case, we show that (the best of multiple randomized executions) of our baseline algorithm meets or exceeds the performance of the short bursts algorithm. Furthermore, our results document the power of our two other new building blocks: an approach that takes a given $r$-way partition of a region (with up to $r=5$ subregions) and finds a new partition that improves the compactness metrics for that set of districts, as well as a local search procedure that uses this same core approach to repeatedly identify additional majority-minority districts within a given $r$-district connected fragment of a statewide plan. Although we have tailored the use of these two elements to our overall framework, we believe that they could lead to additional improvements for other algorithmic approaches to redistricting.

\section{Methodology.}
\subsection{Defining the problem}
The input to our search problem consists of a set of census geographical units that divide a geographical region into building blocks, where for each unit we are given both demographic and geographic data. These units are typically either census tracts, blocks or block groups, and in this paper we use block groups. We let $B$ denote the set of block groups in the region of interest, and we let $p_j$ denote the population of block group $j\in B$. Each block group has an associated (population-weighted) center of gravity, which will be used to compute the Euclidean distance between pairs of block groups. Additionally, there is an underlying adjacency graph $G=(B,E)$ where there is an undirected edge $\{i,j\}\in E$ if block groups $i$ and $j$ share a non-trivial border.

A districting plan for $B$ is a partition of $B$ where each part, or each district, is contiguous according to $G$, meaning the subgraph of $G$ induced by the block groups in the district is connected, and the total population of each district satisfies a population balance constraint. If $D_1,\ldots,D_n\subseteq B$ are the districts in the partition, then this balance constraint is satisfied for district $i\in[n]$ if
\begin{equation}\label{eqn:pop_bal}
    \left|\sum_{j\in D_i}p_j - \hat{p}\ \right| \leq \ve\hat{p},
\end{equation}
where $\hat{p}:=\sum_{j\in B}p_j / n$ is the ``ideal'' district population and $\ve$ is a parameter chosen in advance that dictates how far away district populations can be from the ideal, represented as a proportional difference.

The problem we are trying to solve is to find a districting plan that maximizes the number of districts where a targeted voting age subpopulation constitutes a majority among the total voting age population (VAP). For the remainder of the paper, we will let $T$ denote the targeted subpopulation and refer to districts where they constitute a majority of the voting age population as ``majority-$T$'' districts.

\subsection{Stochastic Hierarchical Partitioning Algorithm}
We adapt the methodology of the stochastic hierarchical partitioning algorithm (SHP) to generate districting plans \cite{fairmandering}. To explain our adaptation, we must first explain the essential elements of the SHP algorithm. At a high level, SHP works by generating a tree to store recursively-defined partitions, where the root represents the region $B$ that we are ultimately trying to partition into districts, and each node in the tree represents a subregion of the root region. The tree is generated by having each node generate multiple partitions of its region into contiguous subregions via a partitioning integer program (PIP) until those subregions satisfy the population constraint (\ref{eqn:pop_bal}) and thus are aptly sized to be considered as potential districts in a districting plan of the root region. Given the set of all such potential districts generated by SHP, we can then find a choice of $n$ mutually-compatible districts that maximizes some chosen objective. In our adaptation, that objective is to maximize the number of majority-$T$ districts.

More formally, a node $(R, s)$ in the SHP tree is a region, $R\subseteq B$, and a ``capacity'', an integer $s$ that represents the number of district-sized subregions (subregions that satisfy the population constraint (\ref{eqn:pop_bal}) that $R$ has the population capacity to create, assuming we could split up the population arbitrarily. In this notation, $(B, n)$ would be the root node for an SHP tree generated to produce a plan with $n$ districts, and $(R, 1)$ would be a leaf node that represents a district given by region $R$. 

To ensure that the representation of $R$ with $s$ districts as given above is implementable, it must be the case that the total population of $R$ is within $\ve s\hat{p}$ from $s\hat{p}$. However, since we are subsequently partitioning this region into $s$ districts, allowing too great a deviation may cause tighter restrictions as we further partition the region. For this reason, the SHP algorithm does the opposite, where the PIP tightens the allowed absolute population deviation from $\ve s\hat{p}$ to $\ve\hat{p}/\lceil\log_2(s)\rceil$, which is tightest at the root and gets more relaxed as we partition nodes deeper in the tree.

The partition process involves randomness to allow for different partitions to be generated each time, but it is also not guaranteed to succeed in finding a partition, even with repeated trials. In the case that it fails to generate a single partition after a certain number of trials, we delete the current node and any siblings, i.e., the other nodes from the partition in which it was created, and trace back up the tree to generate a replacement partition of the old node's parent. 

The split size parameter $z$ controls the number of subregions into which $R$ is split, and the fan out width parameter $w$ controls the targeted number of sample partitions taken for that node. For example, if $z=2$ and $w=3$, then a node will be partitioned up to three times, each of which splits the region into two subregions, resulting in at most three sets of two child nodes. By adjusting the average values of $z$ and $w$, we can adjust the total number of districts produced. Producing more potential districts increases the diversity of options to choose from but also exponentially increases the computation time required to generate the tree. For each node $(R,s)$ in the tree we continue to generate children until $s=1$, and for our experiments, we set $w=3$, and if $s \leq 5$, we set $z=s$, and otherwise set $z=2$. 

The partition process for a node $(R, s)$ can be broken into four steps. First, we determine the capacities of the subregions. If $z=s$, then each subregion must have capacity $1$. If $z=2$, we split $s$ into $s_1$ and $s_2$ by choosing $i\in\{1,2\}$ uniformly at random $s$ times and letting $s_i$ be the number of times $i$ was chosen. Second, we select z samples uniformly at random without replacement from among the block groups in $R$, which will function as ``centers'' of the child subregions $R_1,\ldots,R_z$. Third, if $z=2$, we performing a matching step in which we partition R based on which center each block group is closer to, and we assign the larger of $s_1$ and $s_2$ to the part of the partition with the larger population, and vice versa for the smaller of the two capacities. Fourth and most significantly, we run the PIP to attempt to find an assignment of the block groups in $R$ to the sampled centers. 

The capacity and center sampling steps are random, and one possible outcome is that the PIP is infeasible; in this case we repeat with new samples. Regardless, the results in \cite{fairmandering} suggest that a variety of sampling approaches work equally well, and this method worked well in our case.

In the SHP algorithm, the PIP maximizes a dispersion-based measure of the total compactness of the subregions produced. The PIP has a binary decision variable $x_{ij}\in\{0,1\}$ for each block group $j\in R$ and center $i\in C$ (where $C\subset R$ is the set of centers) that indicates whether $j$ is assigned to the region centered by $i$. The dispersion measure is given by $$
\sum\limits_{i \in C} \sum\limits_{j \in R} (d_{ij})^\alpha p_j x_{ij},$$
where $d_{ij}$ is the Euclidean distance between block group $j$ and center $i$. The distance values are raised to $\alpha\in[1,2]$, which is chosen uniformly at random for each partition to promote more diverse district shapes.

The PIP enforces contiguity of the districts by ensuring that if block group $j$ is assigned to center $i$, then at least one of the block groups neighboring $j$ that are closer than $j$ to $i$ (with respect to the shortest path distance in the subgraph of $G$ induced by $R$) must be assigned to center $i$ as well. This involves computing
\begin{equation}
    S_{ij}:=\{k\mid\{k,j\}\in E_R, d_{ik}^R<d_{ij}^R\},
\end{equation}
which is the set of block groups in $R$ that neighbor $j$ and are closer than $j$ to $i$, where $E_R$ is the set of edges in the subgraph of $G$ induced by $R$ and $d^R_{ij}$ is the shortest path distance between $i$ and $j$ in the same subgraph.

The full PIP is as follows:

\begin{align}
\label{pip:compactness:0}\text{maximize }  & -\displaystyle\sum\limits_{i \in C} \displaystyle\sum\limits_{j \in R} (d_{ij})^\alpha p_j x_{ij}  \\
\label{pip:compactness:1}\text{s.t.} \quad & \displaystyle\sum\limits_{i \in C}  x_{ij}  = 1, \ \ \ \ \ \  \forall j  \in R; \\
& \label{pip:compactness:2}\displaystyle\sum\limits_{j \in R} p_j x_{ij}  \leq  \hat{p} (s_i + \ve / \lceil\log_2(s)\rceil), \ \ \ \ \ \  \forall i \in C; \\
& \label{pip:compactness:3}\displaystyle\sum\limits_{j \in R} p_j x_{ij}  \geq  \hat{p} (s_i - \ve / \lceil\log_2(s)\rceil),  \ \ \ \ \ \  \forall i \in C; \\
& \label{pip:compactness:4} \displaystyle\sum\limits_{k \in S_{ij}}  x_{ik} \geq x_{ij},  \  \ \ \ \ \ \forall i \in C , \  \forall j \in R; \\
& x_{ij} \in \{0,1\}, \ \ \ \ \ \ \forall i \in C , \  \forall j \in R.
\end{align}
Constraint (\ref{pip:compactness:1}) ensures that each block group is assigned to exactly one center, constraints (\ref{pip:compactness:2}) and (\ref{pip:compactness:3}) ensure population balance, and constraint (\ref{pip:compactness:4}) ensures that subregions are contiguous.

\subsection{Majority-\texorpdfstring{$T$}{T} PIP}
In order to generate more majority-$T$ leaf nodes per partition and thus find plans with more majority-$T$ districts, we modify the PIP from the SHP algorithm to maximize not just for compactness of the child subregions but also simultaneously for the number of majority-$T$ subregions generated. For this purpose, we introduce a balance parameter $\beta\in[0,1]$ that specifies the relative importance of these two competing maximization objectives. We will discuss how we choose this parameter in the following section.

To accomplish this, we introduce a binary indicator variable $m_i$ for each center $i$ which is $1$ if the region surrounding center $i$ is majority-$T$ and $0$ otherwise. This can be enforced via a constraint that requires the $T$-population of the region surrounding center $i$ to be at least half the total VAP of the region multiplied by $m_i$. The astute reader might notice that this constraint is quadratic in our current formulation, since it would require $x_{ij}$ to be multiplied by $m_i$, and so one might worry that computational efficiency has been sacrificed. While it is possible to approximate this by a linear constraint that enforces the $T$-population to be at least half an ``ideal'' VAP multiplied by $m_i$ instead, in practice we found that, perhaps surprisingly, the slowdown in an off-the-shelf use of Gurobi with the exact quadratic constraint instead of the approximate linear version was not significant enough to outweigh the improvement in the results obtained.

The new PIP, which we shall call the majority-$T$ PIP, is as follows, where $t_j$ is the $T$-population of block group $j$ and $v_j$ is the VAP of block group $j$. We distinguish between the previous PIP formulation and the new one by including the new parts in red:

\begin{align}\label{pip:compactness+majmin}
    \text{maximize}\quad  & \textcolor{red}{(1-\beta)\sum_{i\in C} m_i } - \textcolor{red}{\beta}\displaystyle\sum\limits_{i \in C} \displaystyle\sum\limits_{j \in R} (d_{ij})^\alpha p_j x_{ij}\\
    \text{s.t.} \quad& \displaystyle\sum\limits_{i \in C}  x_{ij} = 1,  \ \ \ \ \forall j  \in R &\\
    &\displaystyle\sum\limits_{j \in R} p_j x_{ij} \leq  \hat{p} (s_i + \ve / \lceil\log_2(s)\rceil), \ \   \forall i \in C;&\\
    &\displaystyle\sum\limits_{j \in R} p_j x_{ij} \geq  \hat{p} (s_i - \ve / \lceil\log_2(s)\rceil), \ \    \forall i \in C; &\\
    & \displaystyle\sum\limits_{k \in S_{ij}}  x_{ik} \geq x_{ij}, \ \  \forall i \in C , \  \forall j \in R;& \\
    \label{pip:compactness+majmin:5}& \textcolor{red}{\displaystyle\sum\limits_{j\in R} t_jx_{ij} \geq \frac{1}{2}m_i\displaystyle\sum\limits_{j\in R}v_jx_{ij},}\ \ \ \textcolor{red}{\forall i\in C};&\\
    & x_{ij}, \textcolor{red}{m_i}\in \{0,1\},\ \  \forall i \in C, \ \forall j \in R.&
\end{align}

The experiments reported in \cite{fairmandering} indicated that their PIP was effective at partitioning into at most 5 districts. Consequently, one natural strategy is to promote more compact districts by only using the majority-$T$ PIP towards the bottom of the tree. More specifically, when the capacity $s$ of the current node is at or below a certain threshold, which we'll call $s^*$, we set $z=s$ to immediately split the region into leaf nodes and apply the majority-$T$ PIP, and otherwise we use the previous PIP with $z=2$. Increasing $s^*$ encourages a greater fraction of leaf nodes to be majority-$T$ districts, but it also decreases the total number of leaf nodes and thus of feasible plans produced. In our experiments, we set $s^*=5$.

\subsection{\texorpdfstring{$\beta$}{β}-Reoptimization}

There is no clear choice for a value of $\beta$ {\it a priori} for the revised majority-$T$ PIP to properly weight the relative importance of finding compact districts and finding more majority-$T$ districts. However, we do know that if given some node $(R,s)$ in the tree with $s>1$, then for any choice of subregion centers and their associated capacities, any solution obtained when setting $\beta > 0$ cannot give more majority-$T$ districts than a solution obtained when setting $\beta = 0$ and using the same centers and capacities. Furthermore, it is likely that there is some $\beta > 0$ such that the solution obtained has the same number of majority-$T$ districts as that of $\beta = 0$. Thus, we employ a bisection-search-like algorithm, which we call \textit{$\beta$-reoptimization}, to find a solution that matches the number of majority-$T$ districts that is obtained with $\beta = 0$, and places as much weight as possible on the compactness objective.

More specifically, we compute the optimal solution to the majority-$T$ PIP with $\beta=0$ and $\beta=1$ using the same subregion centers and their associated capacities. If the same number of majority-$T$ districts occurs in both, then we use the $\beta=1$ solution in our final plan. Otherwise, we proceed with a standard bisection search (limited to a small fixed number of iterations $b$) to identify the maximum value $\beta^*$ for which the resulting number of majority-$T$ districts matches that of $\beta=0$.

Since $\beta$-reoptimization cannot increase the number of majority-$T$ districts obtained by a given set of centers, there is no reason to use it on nodes whose children are not included as districts in the final plan. Thus, the number of additional PIPs that $\beta$-reoptimization runs is significantly less than $nb$, making it an efficient tool for improving the compactness of solutions generated by our adaptation of SHP.

\subsection{Local Reoptimization}

While our adaptation of SHP is engineered to generate statewide plans with a large number of majority-$T$ districts, there is no guarantee that those plans will have the maximum number achievable. For this reason, we present an algorithm that takes any existing plan and iteratively attempts to find modifications that result in more majority-$T$ districts, taking inspiration from the ReCom MCMC algorithm, but strengthening that approach by replacing randomization with optimization.

The ReCom Markov chain works by treating the state as the current map, and for each iteration, it randomly selects two adjacent districts and randomly redraws the border between those two districts according to a specific procedure. In our algorithm, we also find connected subsets of districts and repartition them, but now we choose a partition by using the majority-$T$ PIP, and we also allow the number of districts in each connected subset to be a small fixed positive integer $r$. Since the aim of our process is to find new partitions (of this subregion) that result in more majority-$T$ districts, we only accept a new partition if it is an improvement, i.e., it increases the number of majority-$T$ districts. Also, instead of choosing a connected subset of districts at random, we enumerate all possible connected subsets of size $r$ and then attempt to find a new partition for those subsets whose priority is deemed high enough with respect to a metric we describe below. Using larger values of $r$ will allow us to find more improvements, but there will be more regions to partition and the individual PIPs will take longer to solve on average. For the sake of overall efficiency, we perform the bulk of our experiments with $r=4$. 

The priority metric to determine the order in which to consider connected subsets of districts of size $r$ is designed to be increasing in the total amount of ``wasted'' majority-$T$ population among the current districts. More specifically, this means computing the proportion of $T$-population in each district and subtracting $1/2$ if that district is majority-$T$ and $0$ otherwise. If all $r$ districts are already majority-$T$, we set the value to be a default of -1. Given $r$ contiguous districts $R_1,\ldots, R_r$, let $P(R_1,\ldots,R_r)$ denote this metric.

It is straightforward to prove the following Lemma.

\begin{lemma}
Given a contiguous region consisting of disjoint subregions $R_1,\ldots,R_r$,
if there exists a different partition of this region with some other set of districts $R_1',\ldots,R_r'$ among which a larger number of districts are majority-$T$, then $P(R_1,\ldots,R_r)>0$.
\end{lemma}

It is useful to note that if all of these $2r$ districts are exactly equal population, then, in fact, the metric must be at least 1/2. Additionally, our algorithm actually implements a slightly stricter version of this metric, namely $P(R_1,\ldots,R_r)-1/2$. The main purpose for this is to make the number of connected subsets of districts that have positive priority significantly smaller and thus feasible to attempt to reoptimize.

After discarding the connected subsets of districts that have negative priority, we attempt to repartition all positive-priority subsets in decreasing order using a priority queue. When an improvement is found, the priority queue is updated by deleting all subsets which used one or more of the districts just successfully repartitioned and then adding the new subsets as given by the new district adjacency graph.

\begin{figure*}
    \centering
    \begin{minipage}{\columnwidth}
        \includegraphics[width=0.95\textwidth]{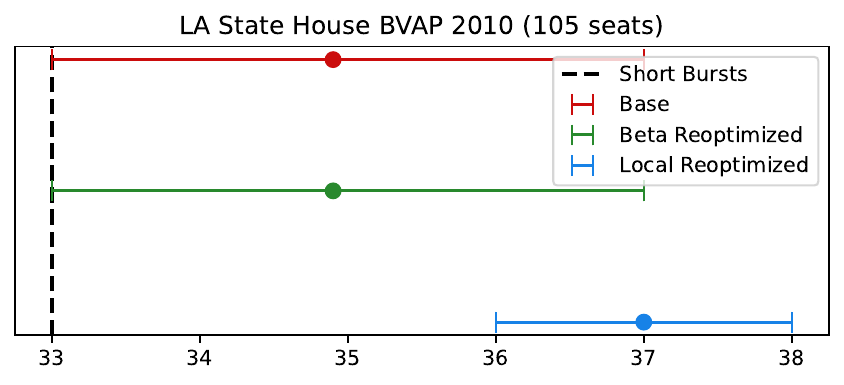}
        \label{subfig:1:whiskers}
    \end{minipage}
    \hfill
    \begin{minipage}{\columnwidth}
        \includegraphics[width=0.95\textwidth]{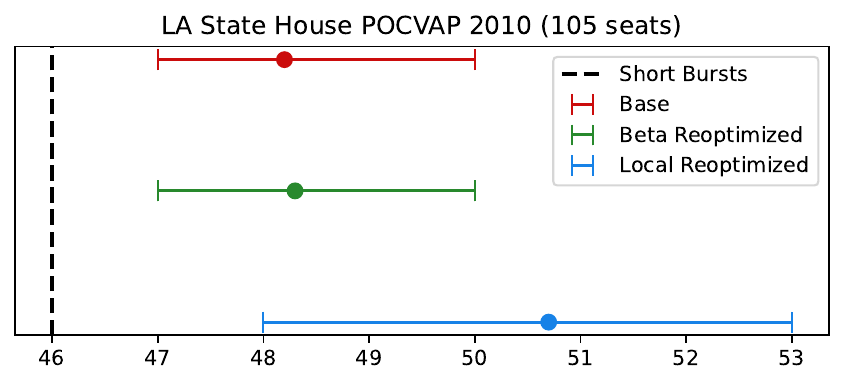}
        \label{subfig:2:whiskers}
    \end{minipage}
    
    \vskip 0pt

    \begin{minipage}{\columnwidth}
        \includegraphics[width=0.95\textwidth]{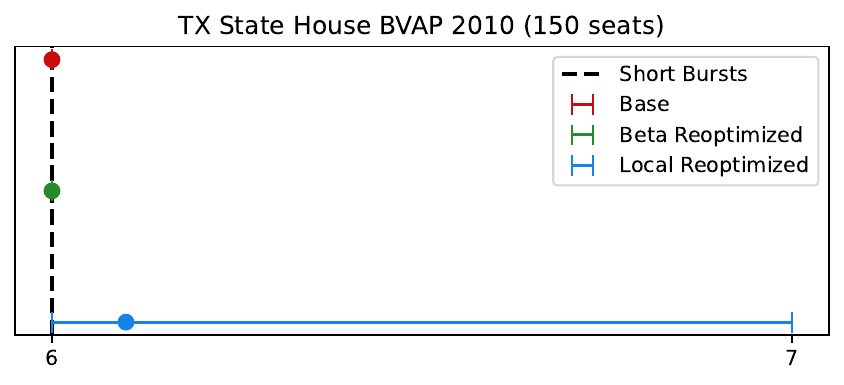}
        \label{subfig:3:whiskers}
    \end{minipage}
    \hfill
    \begin{minipage}{\columnwidth}
        \includegraphics[width=0.95\textwidth]{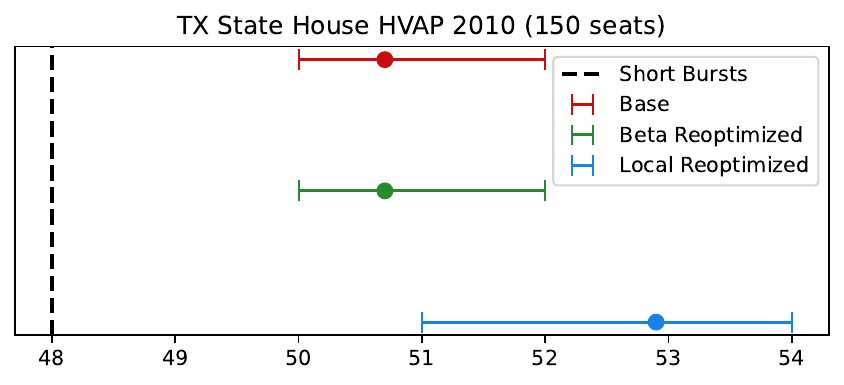}
        \label{subfig:4:whiskers}
    \end{minipage}

    \vskip 0pt

    \begin{minipage}[b]{\columnwidth}
        \includegraphics[width=0.95\textwidth]{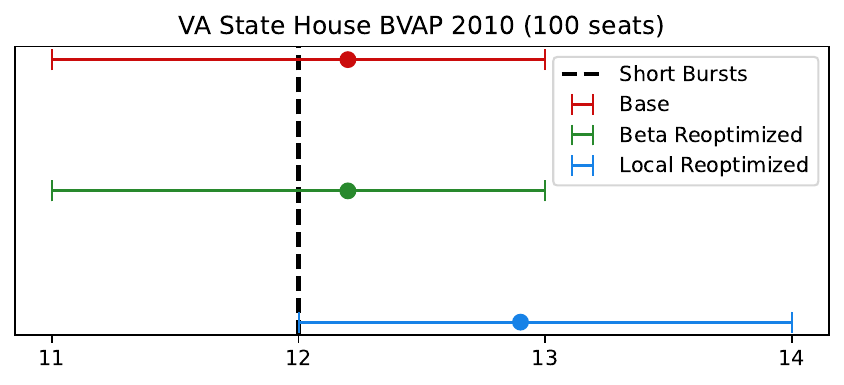}
        \label{subfig:5:whiskers}
    \end{minipage}
    \hfill
    \begin{minipage}[b]{\columnwidth}
        \includegraphics[width=0.95\textwidth]{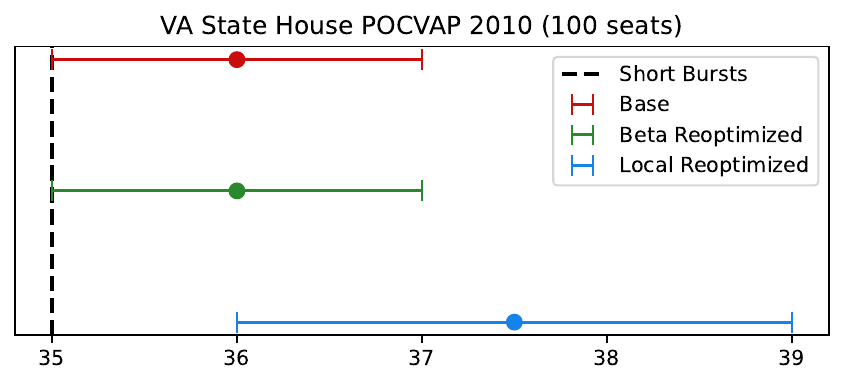}
        \label{subfig:6:whiskers}
    \end{minipage}

    \vskip 0pt

    \begin{minipage}[b]{\columnwidth}
        \includegraphics[width=0.95\textwidth]{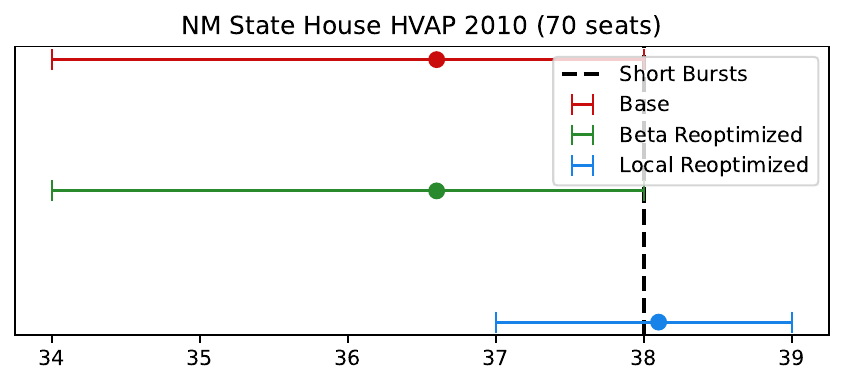}
        \label{subfig:7:whiskers}
    \end{minipage}
    \hfill
    \begin{minipage}[b]{\columnwidth}
        \includegraphics[width=0.95\textwidth]{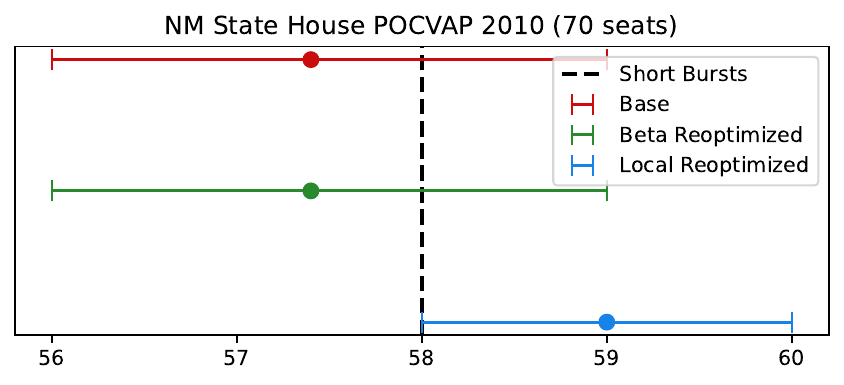}
        \label{subfig:8:whiskers}
    \end{minipage}

    \vskip 0pt
    
    \begin{minipage}{\columnwidth}
        \includegraphics[width=0.95\textwidth]{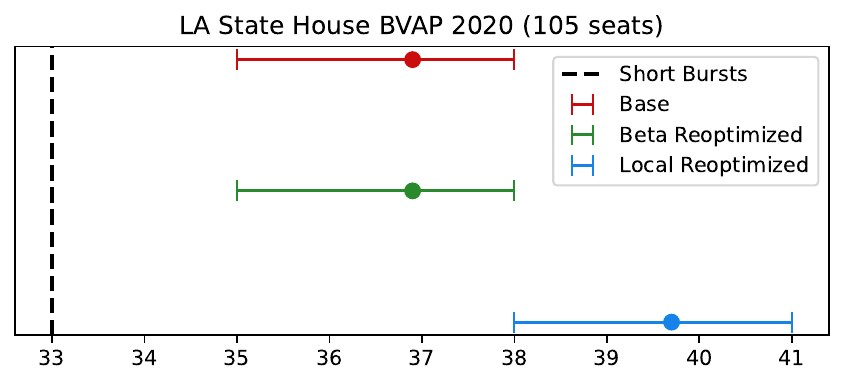}
        \label{subfig:9:whiskers}
    \end{minipage}
    \hfill
    \begin{minipage}{\columnwidth}
        \includegraphics[width=0.95\textwidth]{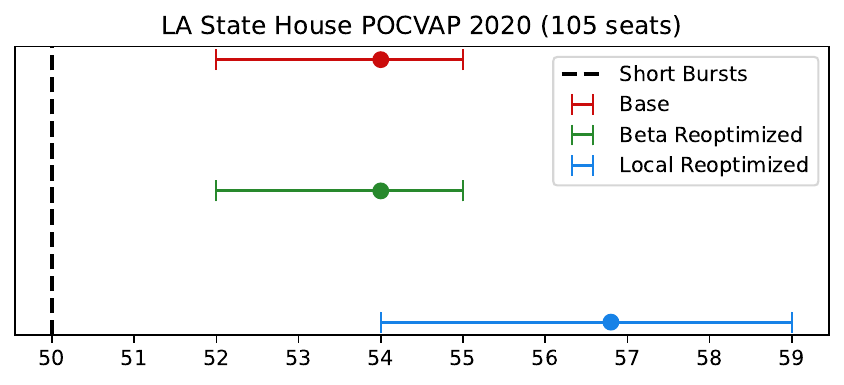}
        \label{subfig:10:whiskers}
    \end{minipage}

    \caption{Numbers of Majority-$T$ districts in each plan generated for several triples of (state, subpopulation $T$, year). For each (state, $T$, year) triple, 10 plans were generated by our adaptation of SHP, and 20 more plans were generated by separately running each plan through $\beta$-reoptimization and local reoptimization. The dots are the averages across each subset of plans, and the bars are the min/max ranges. The x-axis marks the number of majority-$T$ districts, and the dotted line marks the maximum number of majority-$T$ districts found by Cannon et al. \cite{cannon2022votingrightsmarkovchains} }
    \label{fig:whiskers}
\end{figure*}

\subsection{Data}

We obtain the same data used in Cannon et al. \cite{cannon2022votingrightsmarkovchains} for all our experiments, which includes 2010 Census demographic data and Tiger/Line shapefiles, both at the block group level. We download this data via the Census API.

We also obtained 2020 Census demographic data and Tiger/Line shapefiles at the block group level straight from the Census API, and we obtained graphs from Austin Buchanan.

To align with short bursts, we used the same interpretation of BVAP and HVAP as they did, that only counts voting age people who are solely Black or Hispanic and no other race, respectively. For the 2020 data, since we performed our own short bursts experiments, we used the interpretation of BVAP that counts anyone who is Black and is any other number of races as well. This should not affect the conclusions of our experiments, since within each combination of state, year and minority population $T$, our algorithms and the short bursts baseline to which we compare it share the same interpretation of these quantities.

The codebase for this project is publicly available at the following link: \url{https://github.com/D-Brous/gerrypy_daniel}
\section{Results.}

To compare results directly with Cannon et al., we create plans for each of the states considered in their study, namely Louisiana, Texas, Virginia and New Mexico, and for the same minority groups they do in their paper, namely Black voting age population (BVAP), Hispanic voting age population (HVAP), and People of Color voting age population (POCVAP). We also include Louisiana experiments for BVAP and POCVAP using 2020 census data. This was not included in Cannon et al., but we will use this to compare with results obtained from re-running short bursts on that same data. 

For each combination of state, year, and minority population $T$, we create 10 plans using our adaptation of SHP and then run $\beta$-reoptimization on all plans for $b=30$ steps, and we also run local reoptimization with $r=4$ on each of the pre-$\beta$-reoptimized plans. The average run times for these experiments are reported in Table \ref{tab:runtimes}. The run times for $\beta$-reoptimization are not reported since they were small in comparison to the other algorithms. For the two cases of Louisiana 2020 data, we generate 10 random seed plans and perform our own short burst runs on each plan for 500,000 steps with bursts of length 10, which is the highest-scoring burst length in Cannon et al. Figure \ref{fig:whiskers} shows how our algorithm compares to the best plans found by Cannon et al.

\begin{table*}[t]
\centering
\begin{tabular}{|c|c|c|}
    \hline
    & Majority-$T$ SHP & Local Reoptimization\\
    \hline
    LA BVAP 2010 & 0.38 & 0.37 \\
    \hline
    LA POCVAP 2010 & 0.38 & 0.35 \\
    \hline
    TX BVAP 2010 & 5.28 & 1.58 \\
    \hline
    TX HVAP 2010 & 5.56 & 3.97 \\
    \hline
    VA BVAP 2010 & 0.73 & 0.45 \\
    \hline
    VA POCVAP 2010 & 0.87 & 0.59 \\
    \hline
    NM HVAP 2010 & 0.07 & 0.14 \\
    \hline
    NM POCVAP 2010 & 0.07 & 0.06 \\
    \hline
    LA BVAP 2020 & 0.56 & 0.42 \\
    \hline
    LA POCVAP 2020 & 0.56 & 0.45 \\
    \hline
\end{tabular}
\caption{Average run times in hours for majority-$T$ SHP and local reoptimization for each tested combination of state, year, and majority-$T$ population. All experiments were run on a 2023 MacBook Pro with M2 Pro Chip with 16 GB of memory.\label{tab:runtimes}}
\end{table*}

In this figure, the number reported for short bursts in each of the two 2020 plots is the maximum number of majority $T$ districts obtained from the 10 seed plans after 100,000 steps, and not the full 500,000. We reported the former as opposed to the latter since Cannon et al. only ran their algorithm for 100,000 steps. However, the extra steps were intentional, since we aimed to draw a time-normalized comparison by running short bursts on each seed plan for the same amount of time it takes to generate a plan via our adaptation of SHP and then run local reoptimization on that, and 500,000 steps proved to be more than enough. After running short bursts for 5 times as long as in Cannon et al., we found that the maximum number of majority $T$ districts achieved were not significantly affected, increasing by $6.5\%$ on average across the BVAP and POCVAP trials. This demonstrates that increasing computation time has significant diminishing returns, and suggests that short bursts cannot achieve numbers comparable to our algorithms in any reasonable amount of time.

Furthermore, in all of the experiments we conducted, all of our methods tie or outperform short bursts, having more majority-$T$ districts on average, with the only exception being New Mexico, for which only our best run matches the short bursts algorithm. Additionally, local reoptimization increases the average number of majority-$T$ districts significantly in every experiment. While not shown in Figure \ref{fig:whiskers}, we also performed these time-normalized comparison experiments with 2020 Census data for Virginia and New Mexico with the same minority groups as for the 2010 cases and saw similar results, with our algorithms tying or outperforming short bursts while having shorter run times.

We also measure the compactness of the plans generated from each experiment. We use the Polsby Popper compactness measure for this purpose. We chose this metric for two reasons: first, it is one of the most commonly used measures of compactness, and second, since it is rather different from the dispersion metric used in $\beta$-reoptimization, good performance with respect to Polsby Popper suggests that there will be a robustness of performance with respect to any choice of compactness measure. Additionally, the dispersion measure in each PIP's objective function includes a randomly chosen parameter $\alpha\in[1,2]$ for each split, and so it is less clear how to draw a consistent interpretation from that measure. We chose not to employ a Polsby Popper measure in the PIP objective functions due to its non-linearity. 

\begin{figure*}
    \centering
    \begin{minipage}{\columnwidth}
        \includegraphics[width=0.95\textwidth]{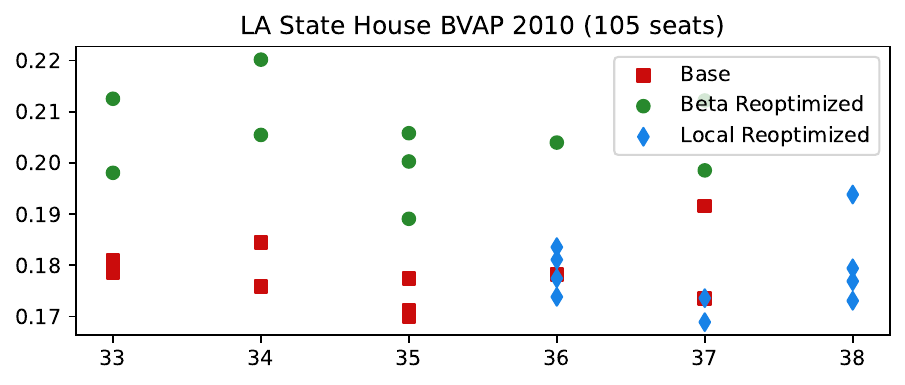}
        \label{subfig:1:scatter}
    \end{minipage}
    \hfill
    \begin{minipage}{\columnwidth}
        \includegraphics[width=0.95\textwidth]{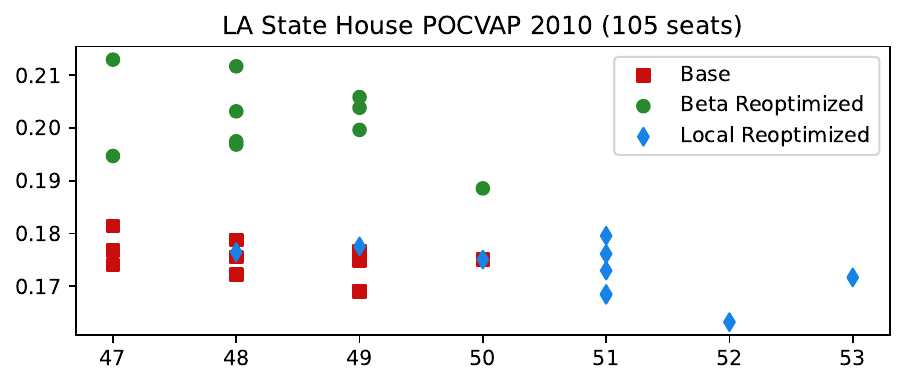}
        \label{subfig:2:scatter}
    \end{minipage}
    
    \vskip 0pt

    \begin{minipage}{\columnwidth}
        \includegraphics[width=0.95\textwidth]{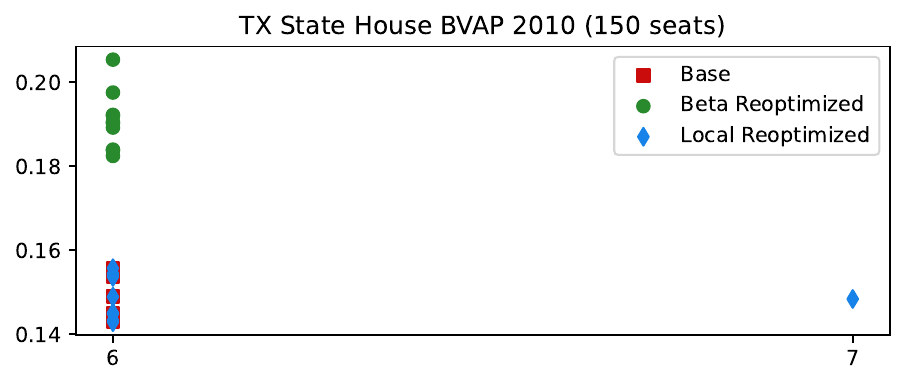}
        \label{subfig:3:scatter}
    \end{minipage}
    \hfill
    \begin{minipage}{\columnwidth}
        \includegraphics[width=0.95\textwidth]{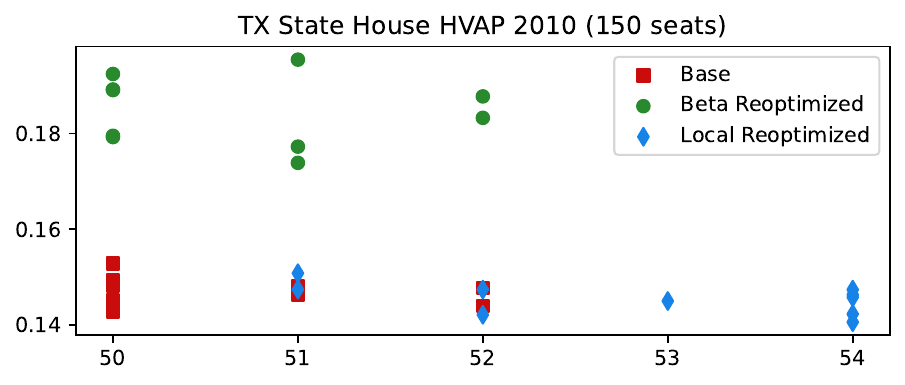}
        \label{subfig:4:scatter}
    \end{minipage}

    \vskip 0pt

    \begin{minipage}[b]{\columnwidth}
        \includegraphics[width=0.95\textwidth]{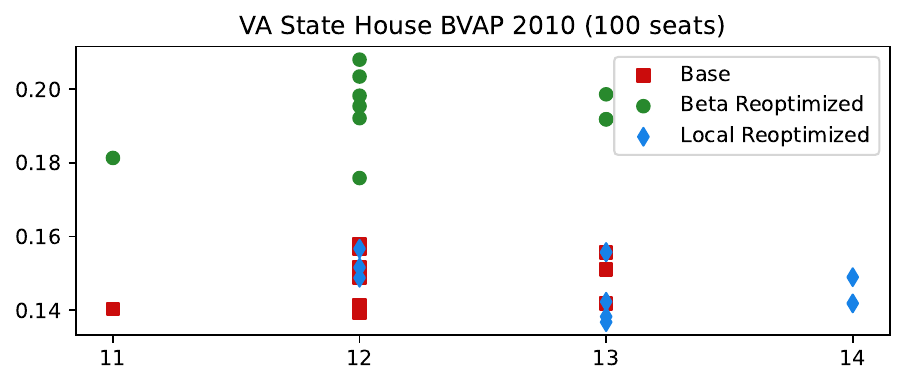}
        \label{subfig:5:scatter}
    \end{minipage}
    \hfill
    \begin{minipage}[b]{\columnwidth}
        \includegraphics[width=0.95\textwidth]{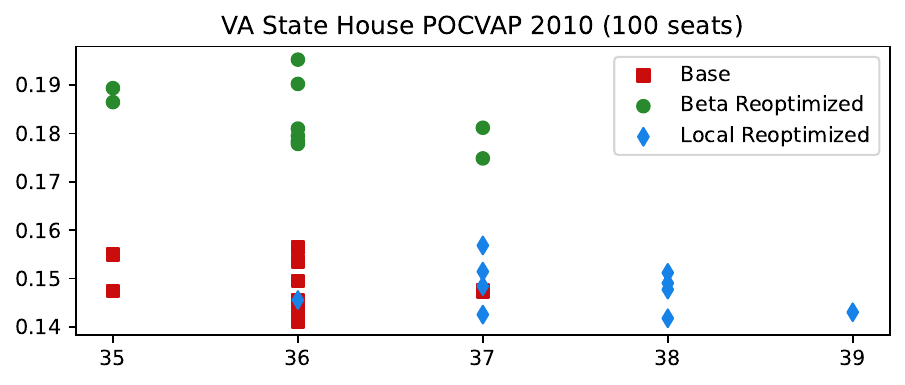}
        \label{subfig:6:scatter}
    \end{minipage}

    \vskip 0pt

    \begin{minipage}[b]{\columnwidth}
        \includegraphics[width=0.95\textwidth]{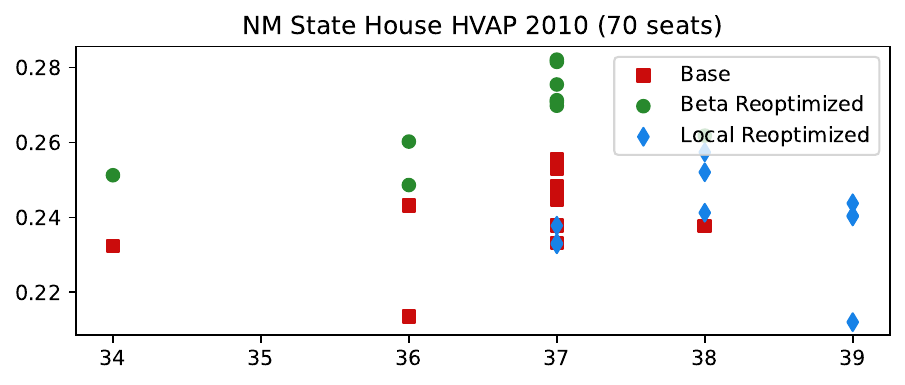}
        \label{subfig:7:scatter}
    \end{minipage}
    \hfill
    \begin{minipage}[b]{\columnwidth}
        \includegraphics[width=0.95\textwidth]{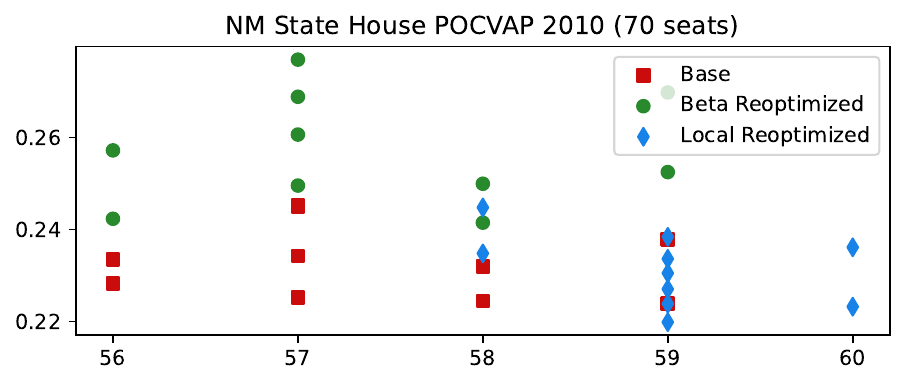}
        \label{subfig:8:scatter}
    \end{minipage}

    \vskip 0pt
    
    \begin{minipage}{\columnwidth}
        \includegraphics[width=0.95\textwidth]{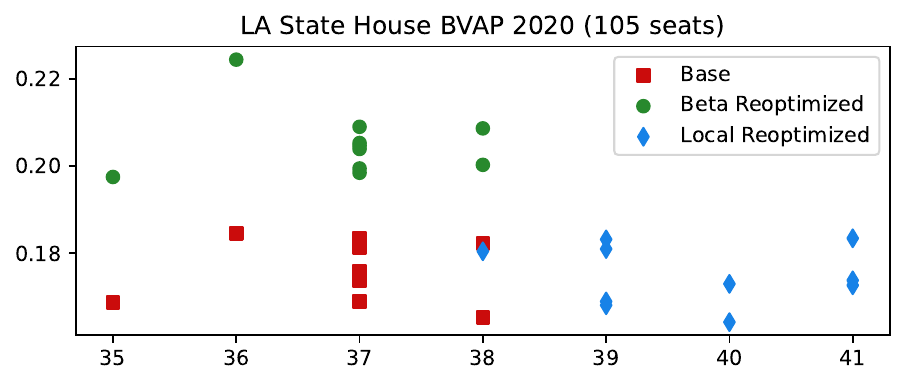}
        \label{subfig:9:scatter}
    \end{minipage}
    \hfill
    \begin{minipage}{\columnwidth}
        \includegraphics[width=0.95\textwidth]{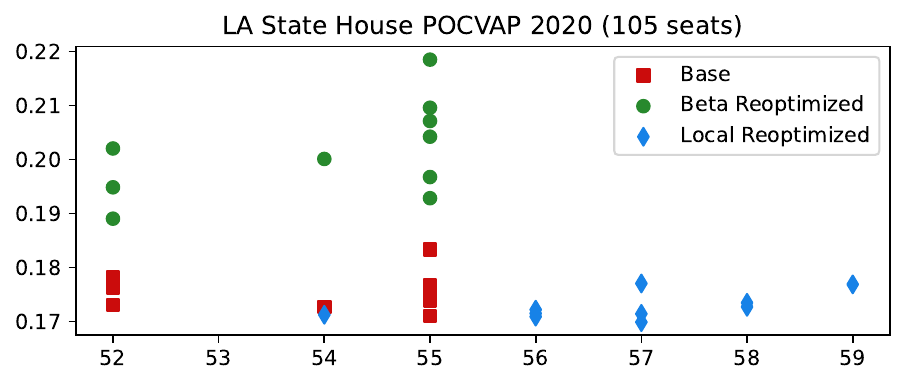}
        \label{subfig:10:scatter}
    \end{minipage}

    \caption{Average Polsby Popper compactness score vs number of majority-$T$ districts in each plan generated for several triples of (state, subpopulation $T$, year). For each (state, $T$, year) triple, 10 plans were generated by our adaptation of SHP, and 20 more plans were generated by separately running each plan through $\beta$-reoptimization and local reoptimization. The x-axis marks the number of majority-$T$ districts in a plan and the y-axis marks the average Polsby Popper score across all the districts in a plan.}  
    \label{fig:scatter}
\end{figure*}

For each plan, we report the average Polsby Popper score across all the districts. Figure \ref{fig:scatter} shows how the average Polsby Popper score varies with the number of majority-$T$ districts for each plan generated.

As expected, $\beta$-reoptimization strictly increases the average compactness scores of each plan generated via our adaptation of SHP, while not affecting the number of majority-$T$ districts. Also, the average compactness score for local reoptimized plans is largely the same as it is for the majority-$T$ SHP plans, which demonstrates that the trade-off between compactness and number of majority-$T$ districts is not exact, having a wide margin for the number of majority-$T$ districts that can be achieved for each level of compactness.

\begin{figure*}
    \centering
    \begin{minipage}[b]{1.1\columnwidth}
        \includegraphics[width=\textwidth]{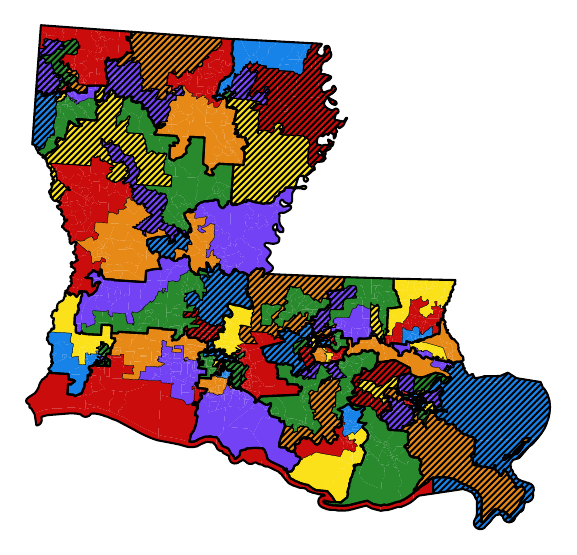}
        \label{subfig:1:LA_plan_beta_reopt_comparison}
    \end{minipage}
    
    \vskip 0pt
    
    \begin{minipage}[b]{1.1\columnwidth}
        \includegraphics[width=\textwidth]{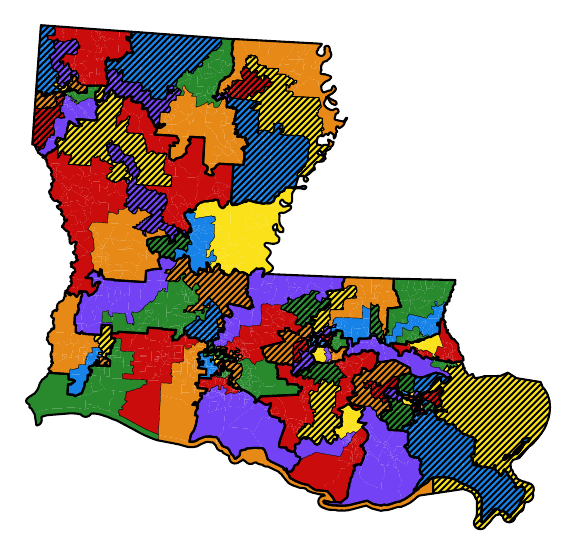}
        \label{subfig:2:LA_plan_beta_reopt_comparison}
    \end{minipage}
    
    \caption{An example plan before and after $\beta$-reoptimization, generated with 2020 data. On the top is LA state house districts optimized for the number of majority-POCVAP districts, but a thick black border is added to display the regions which were reoptimized for compactness via $\beta$-reoptimization. On the bottom is the plan immediately following $\beta$-reoptimization for 30 steps, with the same thick black borders as on the left. Each map has a six coloring on the districts, and the faint white lines within each color zone indicate block group boundaries (may not be visible on print copies). Additionally, every majority-POCVAP district is marked with an overlay of black hash lines.}
    \label{fig:LA_plan_beta_reopt_comparison}
\end{figure*}

\begin{figure}
    \centering
    \begin{minipage}[b]{\columnwidth}
        \includegraphics[width=\textwidth]{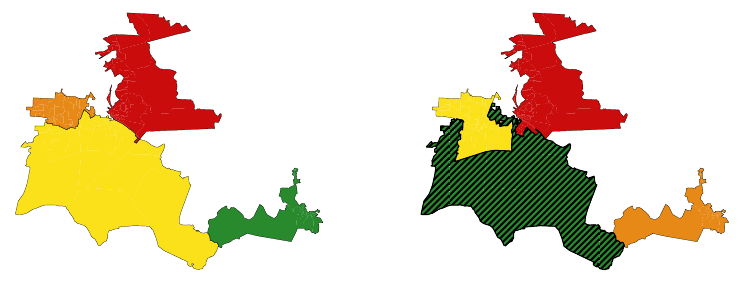}
        \label{subfig:1:LA_plan_local_reopt_steps}
    \end{minipage}
    
    \begin{minipage}[b]{\columnwidth}
        \includegraphics[width=\textwidth]{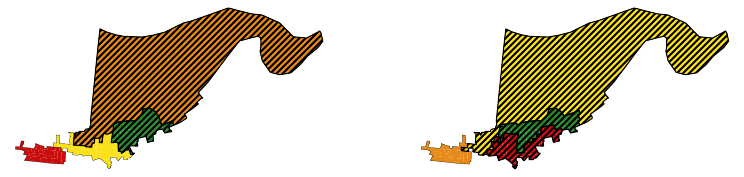}
        \label{subfig:2:LA_plan_local_reopt_steps}
    \end{minipage}

    \begin{minipage}[b]{\columnwidth}
        \includegraphics[width=\textwidth]{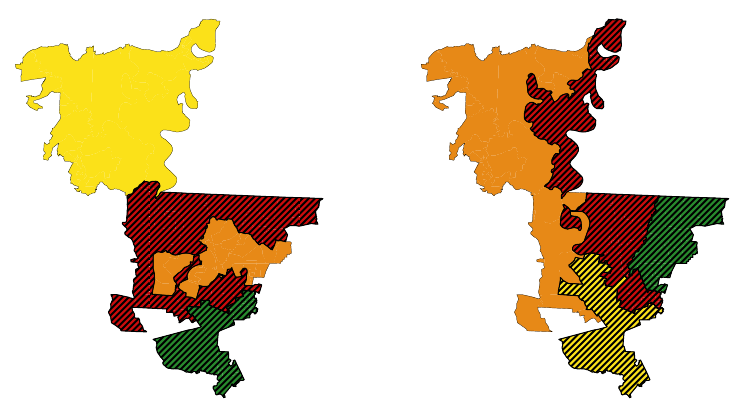}
        \label{subfig:3:LA_plan_local_reopt_steps}
    \end{minipage}

    \begin{minipage}[b]{\columnwidth}
        \includegraphics[width=\textwidth]{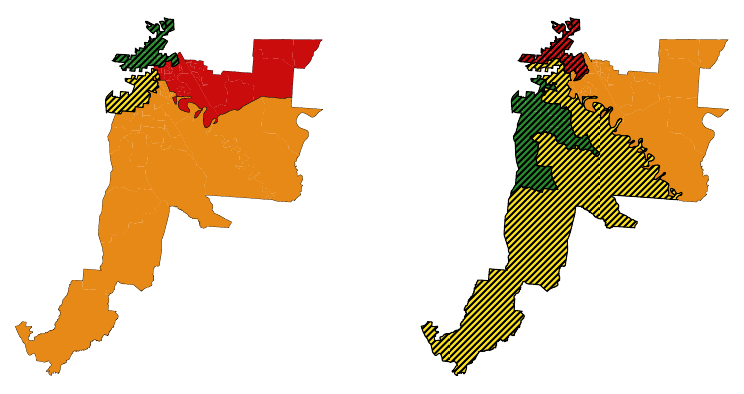}
        \label{subfig:4:LA_plan_local_reopt_steps}
    \end{minipage}
    
    \caption{The regions of 4 districts which were updated in each step of local reoptimization on the LA state house plan shown at the top of Figure \ref{fig:LA_plan_beta_reopt_comparison} Each sub-figure shows the original region on the left and the new region on the right, and the order of steps is top to bottom. Each map has a six coloring on the districts, and the faint white lines within each color zone indicate block group boundaries (may not be visible on print copies). Additionally, every majority-POCVAP district is marked with an overlay of black hash lines.}
    \label{fig:LA_plan_local_reopt_steps}
\end{figure}

In Figure \ref{fig:LA_plan_beta_reopt_comparison}, we show a plan generated via our SHP adaptation, before and after $\beta$-reoptimization, for the Louisiana state house, which has 105 districts, where we maximized for the number of majority-POCVAP districts. This particular plan has 55 majority-POCVAP districts, which is more than the maximum we obtained in our short bursts experiments. We also show the improvement steps in the local reoptimization run for this plan in Figure \ref{fig:LA_plan_local_reopt_steps}. For each of the four improvements found via local reoptimization, the number of majority-POCVAP districts was increased by one.

\section{Conclusions.}

In this paper, we demonstrate that integer programming can be used on both a small and large scale to find districting plans that support a large number of districts in which a targeted subpopulation constitutes a majority. In particular, for all of the data sets considered by the previous work, our randomized methods - based on strong optimization tools - find solutions that have more districts with majority targeted subpopulations than had the current state of the art, which is based on a guided randomized search approach. Furthermore, we expect our results will generalize to all states and for any targeted subpopulation that might constitute a majority in a few districts.

Future work could potentially build on our methodology to provide more general tools that can further explore the Pareto optimal frontier between compactness and numbers of districts in which a targeted subpopulation constitutes a majority. Another potential direction is to add more constraints to the mix, such as reducing the number of county splits, to see how far integer programming can push the frontier of redistricting optimization.

\section{Acknowledgments.}
We would like to thank Austin Buchanan for providing graph data and informing us about different interpretations of BVAP, Sarah Cannon for helping with data questions, and Peter Rock for providing help with the gerrychain package in recreating the short bursts experiments.
\newpage

\bibliography{paper}

\begin{thebibliography}{10}

\bibitem{arredondo2021mathematical}
{\sc V.~Arredondo, M.~Mart{\'i}nez-Panero, T.~Pe{\~{n}}a, and F.~Ricca}, {\em Mathematical political districting taking care of minority groups}, Annals of Operations Research, 305 (2021), pp.~375--402.

\bibitem{becker2022redistricting}
{\sc A.~Becker and J.~Solomon}, {\em Redistricting algorithms}, in Political Geometry: Rethinking Redistricting in the US with Math, Law, and Everything In Between, M.~Duchin and O.~Walch, eds., Birkhäuser Cham, Cham, 2022, ch.~16, pp.~303--340.

\bibitem{belotti2024political}
{\sc P.~Belotti, A.~Buchanan, and S.~Ezazipour}, {\em Political districting to optimize the polsby-popper compactness score with application to voting rights},  (2024).

\bibitem{cannon2022votingrightsmarkovchains}
{\sc S.~Cannon, A.~Goldbloom-Helzner, V.~Gupta, J.~Matthews, and B.~Suwal}, {\em Voting rights, markov chains, and optimization by short bursts},  (2022).

\bibitem{chatterjee2019partisan}
{\sc T.~Chatterjee and B.~DasGupta}, {\em On partisan bias in redistricting: computational complexity meets the science of gerrymandering}, CoRR, abs/1910.01565 (2019).

\bibitem{chestnut}
{\sc {Chestnut~v.~Merrill}}, {\em {U.S. District Court for the Northern District of Alabama}}, 2018.
\newblock Case 2:18-cv-00907-JEO.

\bibitem{deford2022random}
{\sc D.~Deford and M.~Duchin}, {\em Random walks}, in Political Geometry: Rethinking Redistricting in the US with Math, Law, and Everything In Between, M.~Duchin and O.~Walch, eds., Birkhäuser Cham, Cham, 2022, ch.~17, pp.~341--381.

\bibitem{deford2019recombination}
{\sc D.~DeFord, M.~Duchin, and J.~Solomon}, {\em Recombination: A family of {M}arkov chains for redistricting}, arXiv preprint arXiv:1911.05725,  (2019).

\bibitem{duchin2022political}
{\sc M.~Duchin and O.~Walch}, {\em Political Geometry: Rethinking Redistricting in the US with Math, Law, and Everything In Between}, Birkhäuser Cham, Cham, 2022.

\bibitem{dwight}
{\sc {Dwight v. Kemp}}, {\em {U.S. District Court for the Northern District of Georgia}}, 2018.
\newblock Case 1:18-mi-99999-UNA.

\bibitem{garfinkel1970optimal}
{\sc R.~S. Garfinkel and G.~L. Nemhauser}, {\em Optimal political districting by implicit enumeration techniques}, Management Science, 16 (1970), pp.~B--495.

\bibitem{fairmandering}
{\sc W.~Gurnee and D.~B. Shmoys}, {\em Fairmandering: {A} column generation heuristic for fairness-optimized political districting}, in Proceedings of the 2021 {SIAM} Conference on Applied and Computational Discrete Algorithms, {ACDA} 2021, Virtual Conference, July 19-21, 2021, M.~Bender, J.~Gilbert, B.~Hendrickson, and B.~D. Sullivan, eds., {SIAM}, 2021, pp.~88--99.

\bibitem{hess1965nonpartisan}
{\sc S.~W. Hess, J.~Weaver, H.~Siegfeldt, J.~Whelan, and P.~Zitlau}, {\em Nonpartisan political redistricting by computer}, Operations Research, 13 (1965), pp.~998--1006.

\bibitem{johnson}
{\sc {Johnson v. Ardoin}}, {\em {U.S. District Court for the Middle District of Louisiana}}, 2018.
\newblock Case 3:18-cv-00625-SDD-EWD.

\bibitem{kroger2024bounding}
{\sc S.~Kroger, H.~Validi, I.~V. Hicks, and T.~Perini}, {\em Bounding the number and the diameter of optimal compact black-majority districts},  (2024).

\bibitem{kueng2018fair}
{\sc R.~Kueng, D.~G. Mixon, and S.~Villar}, {\em Fair redistricting is hard}, CoRR, abs/1808.08905 (2018).

\bibitem{milligan}
{\sc {Milligan~v.~Merrill}}, {\em {U.S. District Court for the Northern District of Alabama}}, 2021.
\newblock Case 2:21-cv-01530.

\bibitem{negron}
{\sc {Negron v. City of Miami Beach, Florida}}, {\em {United States Court of Appeals, Eleventh Circuit}}, 1997.
\newblock 113 F.3d 1563.

\bibitem{puppe2008computational}
{\sc C.~Puppe and A.~Tasnádi}, {\em A computational approach to unbiased districting}, Mathematical and Computer Modelling, 48 (2008), pp.~1455--1460.
\newblock Mathematical Modeling of Voting Systems and Elections: Theory and Applications.

\bibitem{thornburg}
{\sc {Thornburg~v.~Gingles}}, United States Supreme Court, 478 U.S. 30 (1986).

\end{thebibliography}
\bibliographystyle{siam}

\end{document}